\documentclass[10pt,journal,compsoc]{IEEEtran}
\IEEEoverridecommandlockouts
\usepackage{amsmath,amssymb,amsfonts}
\usepackage{algorithmic}
\usepackage{graphicx}
\usepackage{textcomp}
\usepackage{xcolor}
\usepackage{booktabs}
\usepackage{todonotes}
\usepackage{url}
\usepackage{balance}

\ifCLASSOPTIONcompsoc
  \usepackage[nocompress]{cite}
\else
  \usepackage{cite}
\fi

\usepackage{csquotes}

\usepackage{tikz}				
\usetikzlibrary{shapes,decorations,arrows,calc,arrows.meta,fit,positioning}
\tikzset{
    -Latex,auto,node distance =1 cm and 1 cm,semithick,
    state/.style ={ellipse, draw, minimum width = 0.7 cm},
    point/.style = {circle, draw, inner sep=0.04cm,fill,node contents={}},
    bidirected/.style={Latex-Latex,dashed},
    el/.style = {inner sep=2pt, align=left, sloped}
}

\begin{document}

\title{Code Comprehension Confounders:\\ A Study of Intelligence and Personality}

\author{Stefan Wagner,~\IEEEmembership{Senior Member, IEEE,} Marvin Wyrich%
\IEEEcompsocitemizethanks{\IEEEcompsocthanksitem S.~Wagner and M.~Wyrich are with the
 Institute of Software Engineering,
 University of Stuttgart,
 Stuttgart, Germany.\protect\\
E-mail: \{firstname.lastname\}@iste.uni-stuttgart.de}}

\markboth{}%
{Wagner, Wyrich: How Intelligence and Personality Relate to Code Comprehension Performance}

\IEEEtitleabstractindextext{%
\begin{abstract}
Program comprehension is a cognitive psychological process.
Accordingly, literature and intuition suggest that a developer's intelligence and personality have an impact on their performance in comprehending source code.
Some researchers have made this suggestion in the past when discussing threats to validity of their study results.
However, the lack of studies investigating the relationship of intelligence and personality to performance in code comprehension makes scientifically sound reasoning about their influence difficult.

We conduct the first large-scale empirical evaluation, a correlational study with undergraduates, to investigate the correlation of intelligence and personality with performance in code comprehension, that is, in this context, with correctness in answering comprehension questions on code snippets. 
We found that personality traits are unlikely to impact code comprehension performance, at least not when their influence is considered in isolation.
Conscientiousness, in combination with other factors, however, explains some of the variance in code comprehension performance.
For intelligence, significant small to moderate positive effects on code comprehension performance were found for three of four factors measured, i.e., fluid intelligence, visual perception, and cognitive speed.
Crystallized intelligence has a positive but statistically insignificant effect on code comprehension performance.

According to our results, several intelligence facets as well as the personality trait conscientiousness are potential confounders that should not be neglected in code comprehension studies of individual performance and should be controlled for via an appropriate study design.
We call for the conduct of further studies on the relationship between intelligence and personality with code comprehension, in part because code comprehension involves more facets than we can measure in a single study and because our regression model explains only a small portion of the variance in code comprehension performance.
\end{abstract}

}

\maketitle

\newcommand{\researchQuestionOne}{\textbf{RQ1:} Is there a relationship between code comprehension performance and intelligence?}
\newcommand{\researchQuestionTwo}{\textbf{RQ2:} Is there a relationship between code comprehension performance and specific personality traits?}

\section{Introduction\label{sec:intro}}
Software developers spend more than 50\% of their time on activities related to program comprehension~\cite{Minelli:2015:LastSummer,Xia:2018:Measuring}.
Accordingly, the motivation to investigate the influences on program comprehension with scientific studies is high, for example to work out guidelines for more understandable code or to support developers in understanding code.

Program comprehension is a cognitive psychological process in which, in addition to the characteristics of the code to be understood, the capacities of the person who wants to understand the code play a role.
Accordingly, researchers suspect that intelligence, for example, has an impact on program comprehension performance~\cite{Siegmund:2015:Confounding}.
Since measuring intelligence with validated questionnaires would often exceed the intended time frame of studies, researchers discuss this potential confounding parameter in code comprehension studies as a threat to validity~\cite{Siegmund:2015:Confounding}.
However, these discussions have so far been conducted without a solid understanding of whether intelligence is actually a significant influencing factor on program understanding.
A study on this question is missing so far.

The situation is similar with the personality of developers.
Although personality has not been discussed as a potentially confounding parameter in code comprehension studies yet~\cite{Siegmund:2015:Confounding}, there are several related studies that have shown that different personality traits affect developers' performance in software engineering activities~\cite{Cruz:2015:FortyYearsPersonality, Da:2007:PersoReview, Kanij:2013:Testing, Shoaib:2009:PersoTesting, Wyrich:2019:Theory}.

When designing studies, insights into the influence of intelligence and personality on program comprehension can provide a useful basis for design decisions, which can ultimately lead to greater confidence in the validity of the results and can partially counteract existing uncertainty about what constitutes a good empirical study~\cite{Siegmund:2016:PastPresent,Siegmund:2015:Views}.
For this reason, we conducted the first large-scale empirical evaluation to answer the following research questions:

\begin{itemize}
    \item \researchQuestionOne
    \item \researchQuestionTwo
\end{itemize}

One hundred thirty-five students participated in our study, and the results provide us with initial evidence that intelligence is an influencing factor to be considered, while personality has a less clear impact on code comprehension performance.
At the same time, we emphasize here that a single study cannot fully address such complex cognitive psychological questions from all conceivable perspectives.
Studies that, for example, measure code comprehension as a construct not via correctness but via subjective evaluations, or studies that are conducted with professionals as participants in an industry context, are necessary to complement our results and to obtain a complete picture.

\section{Background and Related Work\label{sec:background}}

Source code understandability is defined as the extent to which ``code possesses the characteristic of understandability to the extent that its purpose is clear to the inspector''~\cite{Boehm:1976:Quantitative}. In this study, we particularly consider bottom-up comprehension, in which a programmer analyses the source code line by line,  deduces \lq chunks\rq{} of higher abstraction from several lines and finally aggregates these chunks into high-level plans~\cite{OBrien:2004:BottomUp}.

When researchers attempt to measure how well a developer has understood a piece of code, they use a variety of measures, such as subjective ratings, the time taken to understand the code or the correctness of answers to comprehension questions~\cite{Oliveira:2020:Evaluating,Baron:2020:Empirical}.
Afterwards, it is possible to analyse the data and, for example, answer specific research questions about the comprehensibility of the code or the ability of developers with certain characteristics to understand code.
In the present study, for example, we provided participants with two different Java methods and measured their code comprehension through the correctness of answers given to comprehension questions. We then measured intelligence and personality using validated questionnaires.
In the end, we were able to analyse the relationship between the comprehension performance and the individual characteristics.

In the design of such experiments, all decisions should be critically reflected to identify if not mitigate as many threats to validity as possible.
Guidelines for conducting experiments, such as those of Wohlin et al.~\cite{Wohlin:2012:Experimentation} can help in this process. Therefore, 
it is positive that the number of program comprehension papers explicitly discussing threats to validity has increased over time~\cite{Schroter:2017:Comprehending}.

One such threat are \textit{confounding variables}.
A confounding variable or \textit{confounder} \enquote{is an extraneous variable whose presence affects the variables being studied so that the results do not reflect the actual relationship between the variables under study}~\cite{Pourhoseingholi:2012:Control}.
For example, in code comprehension experiments the most frequently discussed confounders are the programming experience and familiarity with the study object~\cite{Siegmund:2015:Confounding}.\footnote{Usually, confounders are distinguished from covariates, which are also related to the outcome, but are otherwise not related to the treatment variable. Since the terms are often used interchangeably and the distinction is not relevant for the motivation of our work, we follow~\cite{Siegmund:2015:Confounding} and use confounder as an umbrella term for extraneous variables that are relevant for the interpretation of an observed relationship between two other variables.}
Not identifying and controlling such factors poses a threat to internal validity due to potential false positive errors~\cite{Pourhoseingholi:2012:Control}.

Siegmund and Schumann~\cite{Siegmund:2015:Confounding} conducted a literature survey of papers published between 2001 and 2010 and compiled a list of confounding parameters that are discussed in program comprehension studies.
Combined with a list of possible control techniques, the work is intended to support researchers in producing valid and reliable results.

Their survey results show that 11 studies mentioned intelligence as a confounding parameter, but its influence was analysed only once~\cite{Siegmund:2015:Confounding}.
Although Siegmund and Schumann identified other individual background variables, such as culture and gender, as potential confounding parameters, personality is not part of the list.
Our study aims to provide specific evidence whether the list should be extended to include the variable personality and whether intelligence actually has the presumed impact on performance in code understanding.

We investigate the influence of intelligence and personality in a joint study because both are individual characteristics that share similar traits, such as remaining relatively stable over a lifetime~\cite{Deary:2000:Stability,Caspi:2000:Child}, yet represent distinct constructs.
In addition, their joint measurement integrates well into our planned study design, ultimately saving resources for us and potential participants by not conducting multiple separate studies.

\subsection{Intelligence in Software Engineering}

\begin{figure*}[tbp]
    \centering
    \includegraphics[width=1\textwidth]{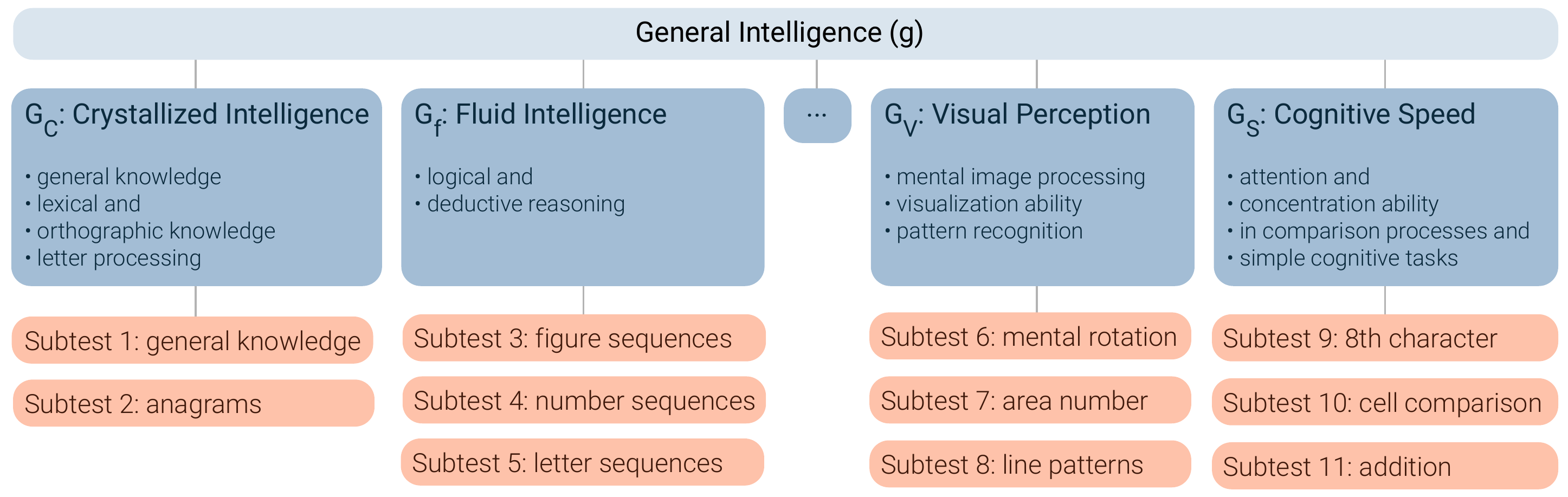}
    \caption{Simplified representation of \textit{LPS-2} for measuring intelligence, depicted in the structure of the three-stratum model by Carroll~\cite{Carroll:1993:Human,Carroll:2005:ThreeStratum}. Figure based on~\cite{lps2}. Three dots in the middle indicate the presence of other second-order factors that determine general intelligence but are not measured by the LPS-2, i.e., memory and learning, auditory perception, retrieval ability, and processing speed.}
    \label{fig:threestratum}
\end{figure*}

In the context of our study, intelligence as a construct is defined by the test used to measure it~\cite{lps2}, and we elaborate on the test further in sections~\ref{sec:materials} and~\ref{methods_variables}.
At this point, it is sufficient to understand that our assumptions (and those of the test) are based on John Caroll's Three-Stratum Theory of intelligence~\cite{Carroll:1993:Human,Carroll:2005:ThreeStratum}.
A series of subtests administered to participants in our study operationalizes first-order factors and allows the measurement of four second-order factors, i.e., crystallized intelligence, fluid intelligence, visual perception, and cognitive speed.
All test performances combined serve to estimate general intelligence, also called g factor, as a third-order factor.
Fig.~\ref{fig:threestratum} visualizes the used intelligence test in the structure of the three-stratum theory.

When researchers refer to intelligence as a potential confounding parameter in discussions of their code comprehension studies~\cite{Siegmund:2015:Confounding} and do not further specify intelligence as a construct, we suspect that this is deliberately left open. 
Whether the g factor, certain dimensions of intelligence, or a combination of both could be responsible for some developers understanding code more easily than others is difficult to assume as long as we still know too little about the concrete cognitive processes in the brain of a developer or even concrete studies on the research question, such as the present one, exist.

Fortunately, in recent years there has been a trend towards more neuroscience and physiological studies in the field of program comprehension~\cite{Siegmund:2020:Crazy,Fakhoury:2018:Objective,Peitek:2018:Simultaneous}, for example, using fMRI scanners to investigate the active brain regions and cognitive processes of developers during source code comprehension.
Peitek et al.~\cite{Peitek:2018:Look} found in an fMRI study with 17 participants and subsequent replication with 11 participants that bottom-up program comprehension involves activation of five brain regions that are related to working memory, attention, and language processing.
Accordingly, we suspect that at least the intelligence factors cognitive speed and crystallized intelligence should also correlate positively with performance in code comprehension in our study.

While such studies with new ideas for measurement methods already provide us with interesting early research findings, their implementation currently still represents an evaluation of these very measurement methods.
To the best of our knowledge, there is no large-scale study that is directly aimed at investigating the relationship between program comprehension performance and intelligence.
However, there are a few studies in software engineering which measured intelligence as part of their design.

Ko and Uttl~\cite{Ko:2003:Individual} conducted an exploratory experiment with 75 undergraduates and measured, among other individual characteristics, verbal intelligence with the \textit{Vocab27} test as well as problem-solving ability with a \textit{problem-solving} test consisting of items from various intelligence tests.
These individual differences did not appear to correlate with the success in debugging a program in an unfamiliar programming system.

Mindermann and Wagner~\cite{Mindermann:2020:Fluid} found in an experiment with 76 undergraduates that the successful usage of cryptographic libraries in terms of effectiveness, efficiency and satisfaction with the help of examples is not related to the participants' fluid intelligence.
Fluid intelligence can be summarized by the ability for logical and deductive reasoning, which is why the non-existent influence on effectiveness and efficiency of task processing were surprising for the authors of the study.

\subsection{Personality in Software Engineering}

Since there are many definitions of the term \textit{personality}, we are guided by what Ryckman describes as a consensus among investigators, that is, \enquote{the dynamic and organized set of characteristics possessed by a person that uniquely influences his or her cognitions, motivations, and behaviors in various situations}~\cite{Ryckman:2012:Personality}.

This definition was also the basis for a systematic mapping study by Cruz et al.~\cite{Cruz:2015:FortyYearsPersonality}. 
Their study provides us with valuable insights into forty years of research on personality in software engineering (1970--2010).
Several studies found that personality traits correlate with performance in various SE tasks~\cite{Cruz:2015:FortyYearsPersonality, Da:2007:PersoReview, Kanij:2013:Testing, Shoaib:2009:PersoTesting, Wyrich:2019:Theory}, others found no significant relationship between personality and programming performance~\cite[e.g.]{Bell:2010:Personality}.
However, of the nine papers found in the mapping study that could be classified under the research topic of individual performance, none explicitly examined the influence of personality on code comprehension performance~\cite{Cruz:2015:FortyYearsPersonality}.

Karimi et al.~\cite{Karimi:2016:Links} investigated how personality affects programming styles, including the approach to code understanding.
Programmers with high conscientiousness tended to use depth-first style, and those high in openness to experience tended to use a breadth-first style.
They further found that programmers who tended to use a depth-first approach often showed better programming performance.

The closest to our research questions is a study by Arockiam et al.~\cite{Arockiam:2005:ObjectPersonality} on the influence of personality traits on the correctness of comprehension questions on C++ programs.
Unfortunately, the paper lacks a comprehensive description of the design that would enhance our confidence in the validity of the results.
Furthermore, the used personality model and test seem not to be validated.

In summary, neither the influence of personality nor that of intelligence on code comprehension performance has been sufficiently studied to date.
Since such insights would be useful for researchers in the field of code comprehension in designing studies and ensuring validity, we are taking a step in this direction and are beginning to fill the identified research gap.

\section{Methodology\label{sec:methodology}}

\begin{figure*}[tbp]
    \centering
    \includegraphics[width=1\textwidth]{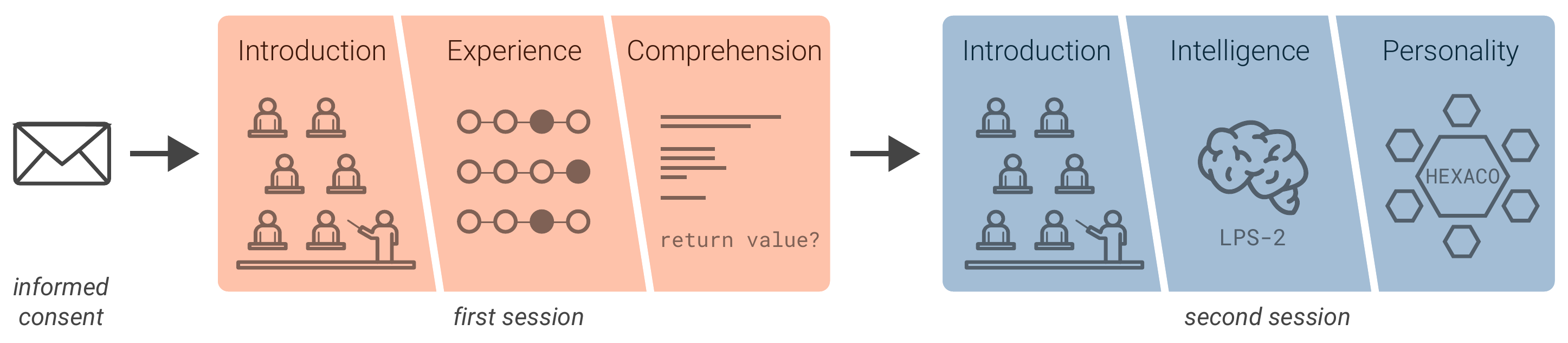}
    \caption{Schematic representation of the research design. Icons from~\cite{NounProject:2021:DairyFree}}
    \label{fig:procedure}
\end{figure*}

We follow the guidelines of Jedlitschka et al.~\cite{Jedlitschka:2008:Reporting} on reporting experiments in software engineering.

\subsection{Goals}

The goal of the study is to analyse the impact of different facets of intelligence as well as personality traits on the performance in understanding source code.
To this end, we formulated the following two research questions:

\begin{itemize}
    \item \researchQuestionOne
    \item \researchQuestionTwo
\end{itemize}

\subsection{Research Design} %

We conducted a correlational study: We did not manipulate intelligence or personality, but measured them as given subject characteristics.
This is a common study design in any study where the independent variable cannot be manipulated for practical or ethical reasons~\cite{Cronbach:1957:TwoDisciplines,Rohrer:2018:ThinkingClearly}.
With this in mind, in section~\ref{methods_causal} we describe a causal model that provides the basis for a discussion of causal inferences from our collected data~\cite{Rohrer:2018:ThinkingClearly}.

We provided participants with a consent form and informed them in writing about the aim of the study.
The study took place simultaneously and on-site for all participants at two given time slots on two days.
Participants had the opportunity to ask questions at the beginning of each session.

In the first session, the participants received a short questionnaire on their programming experience as well as two code snippets for which they had to answer comprehension questions under time limits.
In the second session, participants took an intelligence test and a personality test.
We linked the data of both sessions by a panel code\footnote{Participants answered 10 questions, with each answer consisting of one to a maximum of two characters. The concatenation of the characters results in the panel code. For example, one question is: \textit{Your birthplace, 1st and 2nd letter}. We only considered data for which there is a matching panel code in both sessions in the analysis.} that preserved the anonymity of the participants.

A schematic representation of the research design is provided in \figurename~\ref{fig:procedure}.

\subsection{Participants}

We consider everyone with at least one year of Java programming experience suitable for the experiment.
Therefore, we invited a convenience sample of students of one of our computer science programs (Computer Science, Software Engineering, Media Informatics, Data Science) in their second year to participate in our study.
All had mandatory Java courses in their first year.
We see the sample properties of interest, namely enough experience to comprehend medium to understand Java code, ensured by our sampling strategy so that the findings can be partially transferred to a population of experienced Java software engineers.
We will discuss limitations of our sampling strategy and their implications in section~\ref{sec:limitations}.

We recruited students enrolled in the same course to keep their academic experience level and programming language experience similar.
As part of their study duties, students had to participate in any study offered by the faculty.
Students had the right to register for a study and then withdraw their participation at any time without consequences.

We informed the participants in writing before the first session about the study design, health risks, privacy and ethical issues and our contact details.
Furthermore, all goals were open, and the study contains no deceptions of the participants.
The panel code allowed the organizers to provide the individual's results for their personality and intelligence tests.
This aspect was intended to motivate participation in the study.

To determine the needed sample size for our analyses, we conducted an a-priori power analysis. By convention, we used $\alpha=.05$ and $\beta=.2$. We wanted to be able to detect small effect sizes because they could still be interesting as confounding factors in a comprehension study. Therefore, we chose what Cohen \cite{cohen1992power} considers a small effect: 0.2. Using the \textit{pwr.t.test} function of the \textit{pwr} R package, we calculated an optimal sample size of 198.

\subsection{Tasks}

Participants were shown two independent Java methods, one after the other.
For each code snippet, five input values were given, for which the participants were asked to specify the return values according to the Javadoc and to determine the actual return value.
Participants knew that the answers would be rated on correctness.

Since we told our participants that there might be bugs in the code, they could not rely on the Javadoc comment and had to understand what the code actually does.
We consider the deviation of the documentation from the code and the inspection based on concrete values for the input parameters to be a realistic scenario.
Furthermore, the task is in line with the conceptual model that a developer in a maintenance scenario iteratively constructs and tests hypotheses about the functioning of the code during program understanding~\cite{Mayrhauser:1995:ProgramComp}.

There was a time limit of 12 minutes for processing each of the two code snippets.
In between, there was a short break of two minutes.

\subsection{Experimental Materials}
\label{sec:materials}

The experiment took place in both sessions in a large lecture hall.
Neither the participants nor the experimenters used electronic devices.
All materials were presented to the participants on paper.
We make all experimental materials publicly available (see Section~\ref{sec:data}) except for the personality and intelligence tests, which we cannot republish for legal reasons.

\subsubsection{Code Snippets}

We used a total of two Java code snippets to conduct the study.
Each code snippet consisted of exactly one method and its Javadoc documentation.
The code was highlighted as in an Eclipse IDE with default settings.

The first code snippet to be understood was a solution to a coding challenge~\cite{Wyrich:2019:Theory}, such as those given in programming contests or technical interviews.
The method comprises 15 SLOC, has two parameters and contains, among other things, two nested for-loops.

The second code snippet was a method from the Apache Commons library for converting an integer to a boolean object.
The method spans 18 SLOC, has four parameters, and is characterized by several if-else branches.

We selected the snippets in a way that no uncommon prior knowledge on, e.g., frameworks, would be required to understand them.
As a result, the code contained mostly primitive data types and the features of newer Java versions were avoided.
Cognitive complexity, a validated metric for assessing the comprehensibility of methods~\cite{Baron:2020:Empirical}, was 9 for the first and 8 for the second method, corresponding to moderately difficult comprehensibility.

In our study, we considered two code snippets to be a sufficient number to maintain a balance between appropriate time commitment from participants and generalizability of results.
First, understanding two code snippets means that participants already have to concentrate for 24 minutes, and second, from an ethical point of view, no more time is taken up by participants than is probably necessary to gain insightful knowledge.
Since we have limited ourselves in the selection of code snippets as previously described, we view the two selected snippets as representative of code with the previously described characteristics.
We acknowledge that this is a personal view of the authors, and different studies have handled the selection and number of code snippets very differently so far.

\subsubsection{Comprehension Questions}

For each of the two snippets, we provided the participants  with a paper-based form which included five rows of a three-column table that had to be filled in.
The cells of the first column each contained a method call, for example \verb toBooleanObject("1,1,0,null") .
The other two columns had to be filled with the actual return value of the method and the expected return value according to the Javadoc.

\subsubsection{Questionnaires}

To measure programming experience, we followed the recommendations of Siegmund et al.~\cite{Siegmund:2014:MeasuringExp} to use self-assessment questions on general programming experience, programming experience compared to fellow students and with the object-oriented paradigm.

For the measurement of intelligence, we searched for an intelligence test that is established in psychology with a correspondingly good validation. Furthermore, the test should be detailed enough to distinguish between different intelligence factors, but also short enough to be conducted together with the personality test in a lecture slot of 90 minutes. The latter was important to be able to have the participants on-site and in an available lecture hall. Furthermore, it should be a paper test, as not all participants might carry a suitable device for an electronic test. Ideally, the tests should be freely available to support the easy replication of our study. 

Unfortunately, we found only the LPS-2 questionnaire~\cite{lps2} fulfils our other criteria but needs to be bought from the publisher. LPS-2 measures four different factors of intelligence: crystallized intelligence (general knowledge, lexical and orthographic knowledge), fluid intelligence (logical and deductive reasoning), visual perception (visualization capability, pattern recognition), and cognitive speed (e.g., ability to concentrate in simple cognitive tasks).
These four factors are operationalized by eleven subtests whose net processing time is 39 minutes. In total, about one hour should be scheduled to conduct the test.
Fig.~\ref{fig:threestratum} shows how the 11 subtests are related to the factors of intelligence.
The test was validated with 2,583 participants~\cite{lps2}, showing an internal consistency in all four subfactors between .86 and .94 with .96 for the general intelligence score in the form we used. Construct validity was shown using confirmatory factor analysis. Regarding criterion validity, its results corresponded well with other intelligence tests.

The scores of the subtests are summed up regarding the four factors and mapped to age-adjusted IQ values within a 95\% CI.
Scoring the intelligence test is done using a set of templates and takes a few minutes per participant.
The total score for all subtests represents an estimate of overall intellectual capacity, commonly referred to as general intelligence (g).

Similarly, as for intelligence, we looked for a validated personality test accepted in psychology that can be done on paper in the same time slot as the intelligence test. ``There is little doubt that the Five-Factor Model (FFM) of personality traits (the ‘Big Five’) is currently the dominant paradigm in personality research, and one of the most influential models in all of psychology.''~\cite{McCrae:2020} We chose the validated German version of the 100 item HEXACO Personality Inventory-Revise (PI-R)~\cite{Lee:2018:Psychometric} which is a variation that adds a sixth dimension. Lee and Ashton have empirical support for this sixth factor
from principal component analysis.
The questionnaire contains 100 statements on which a participant must self-assess on a scale from fully agree to fully disagree.
HEXACO assesses six major dimensions of personality: \textit{H}onesty-Humility, \textit{E}motionality, E\textit{x}traversion, \textit{A}greeableness, \textit{C}onscientiousness and \textit{O}penness to Experience~\cite{Ashton:2001:Theoretical}.
In a large validation study~\cite{Lee:2018:Psychometric}, the facets in the self-assessment had a mean reliability (internal consistency) of about $\alpha = .70$. Furthermore, principal component analyses supported the chosen facets that have low inter-correlations. Furthermore, the test is freely available~\cite{Ashton:2001:Theoretical}.

\subsection{Hypotheses, Parameters, and Variables\label{methods_variables}}

The variables relevant to RQ1 are code comprehension performance and the four factors of intelligence mentioned in~\ref{sec:materials}, i.e., crystallized intelligence ($G_c$), fluid intelligence ($G_f$), visual perception ($G_v$) and cognitive speed ($G_s$).
The average IQ values of university students for the four factors are each in the range of 100 to 107 with a standard deviation between 13 and 15~\cite{lps2}.
We expect all factors to have a positive impact on code comprehension performance.
For $G_f$ we assume the highest correlation, since code comprehension intuitively has a lot to do with logical thinking.
Similarly, we argue for the positive impact of $G_s$, a measure of attention and concentration ability in simple cognitive tasks.
Visual perception ($G_v$) ability should also show a positive correlation in the data for code that represents a visually processed form of knowledge.
We assume the weakest positive impact for $G_c$, on which, for example, knowledge of the Java programming language can be mapped, which is in principle relevant for code comprehension, but also only up to a certain degree, determined and limited by the requirements of the tasks.

We will address the statistical relationship to general intelligence exploratively in the results and discussion and will not formulate a hypothesis.
The construct of general intelligence is less accessible than the four factors of intelligence mentioned above, which are, first, well-defined and psychometrically validated~\cite{lps2} and, second, allow for a finer-grained analysis of specific influences on code comprehension.
It is believed that general intelligence has its own impact on performance and is positively correlated with the four factors\cite{lps2,Bickley:1995:ThreeStratum}.
To also simplify the causal model in relation to mediating variables~\cite{Rohrer:2018:ThinkingClearly}, we therefore omit general intelligence and do not consider it for the hypothesis analysis.

Code comprehension is measured by the correctness of answers to comprehension questions on two independent code snippets.
Correctness is one of the most commonly used measures to assess how well a participant understood source code~\cite{Oliveira:2020:Evaluating,Baron:2020:Empirical}, and depending on the response format, the measurement is reliable even in experiments conducted synchronously with many participants.
For each of the two code snippets, participants could give 10 answers, each of which was scored with one point for a correct answer.
In the analysis we noticed that for the second snippet a question regarding the return value according to Javadoc could not be answered unambiguously, and therefore we decided not to score the answers to this subtask.
Accordingly, code comprehension performance ranges from 0 to 19. We deliberately refrained from using individual measurements of time for the comprehension tasks as this would not have been practical in the lecture hall setting and instead introduced a general time limit.\\

\noindent
$H_1$: Crystallized intelligence ($G_c$) is positively correlated with code comprehension performance.\\
$H_2$: Fluid intelligence ($G_f$) is positively correlated with code comprehension performance.\\
$H_3$: Visual perception ($G_v$) is positively correlated with code comprehension performance.\\
$H_4$: Cognitive speed ($G_s$) is positively correlated with code comprehension performance.\\

To answer RQ2, personality was operationalized by the six dimensions of the HEXACO personality model.
In hypothesizing, we limited ourselves to two traits for which we could find the most evidence of possible impact on performance in scientific literature, that is, conscientiousness ($\mathrm{range}=[1,5]$) and openness to experience ($r=[1,5]$).

Both personality traits were found in previous studies to be associated with performance in various software engineering activities.
We suspect a positive correlation for both with code comprehension performance, since persons with high values for these personality traits have argumentatively good prerequisites to also successfully work on the comprehension tasks.
People with high values for conscientiousness \enquote{organize their time and their physical surroundings, work in a disciplined way toward their goals, strive for accuracy and perfection in their tasks, and deliberate carefully when making decisions}~\cite{HEXACO:website}.
Persons with low values for openness to experience tend to \enquote{feel little intellectual curiosity, avoid creative pursuits}~\cite{HEXACO:website}.\\

\noindent $H_5$: Conscientiousness is positively correlated with code comprehension performance.\\
\noindent $H_6$: Openness to Experience is positively correlated with code comprehension performance.\\

Programming experience was rated following the recommendation in \cite{Siegmund:2014:MeasuringExp} on three 10-point scales from \emph{very inexperienced} to \emph{very experienced}.
Experience is known as a potential covariate and was therefore measured to control for it by subsequent analysis of its influence on code comprehension performance~\cite{Siegmund:2015:Confounding}.

\subsection{Causal Inferences\label{methods_causal}}

We have already mentioned that this is a correlational study, and we could elaborate at this point on why appropriate caution is needed in interpreting potential correlations, that they are only cause-effect relationships with some probability, not guaranteed.
One issue is that \enquote{carefully crafted language will not prevent readers--let alone the public--from jumping to causal conclusions}~\cite{Rohrer:2018:ThinkingClearly}.
Moreover, we strive to close the gap between observational data and causal conclusions as good as we can, to counteract the issue of uncertainty in the design of empirical studies motivated at the beginning with the greatest possible certainty about the concrete nature of influence of our independent variables.

\begin{figure}
  \begin{center}
  \sffamily
  \footnotesize
    \begin{tikzpicture}
        \node (6) [text=gray] {General Intelligence};
        \node (2) [below left = 1cm and -0.5cm of 6, align=center] {Fluid\\Intelligence};
        \node (1) [left = 0.5cm of 2, align=center] {Crystallized\\Intelligence};
        \node (3) [below right = 1cm and -0.5cm of 6, align=center] {Visual\\Perception};
        \node (5) [right = 0.5cm of 3, align=center]{Cognitive\\Speed};
        
        \node (4) [below = 3cm of 6] {Code Comprehension Performance};
        
        \node (7) [below = 1cm of 4]{Conscientiousness};
        \node (8) [right = of 7, align=center]{Openness to\\Experience};
        \node (9) [text=gray, left = of 7, align=center]{Programming\\Experience};
    
        \path [gray] (6) edge (4);
        \path (3) edge (4);
        \path (2) edge (4);
        \path (1) edge (4);
        \path (5) edge (4);
        \path [gray] (1) edge (6);
        \path [gray] (2) edge (6);
        \path [gray] (3) edge (6);
        \path [gray] (5) edge (6);
        \path (7) edge (4);
        \path (8) edge (4);
        \path [gray] (9) edge (4);
    \end{tikzpicture}
  \end{center}
  \caption{Causal diagram of the experiment variables}
  \label{fig:causal-diagram}
\end{figure}
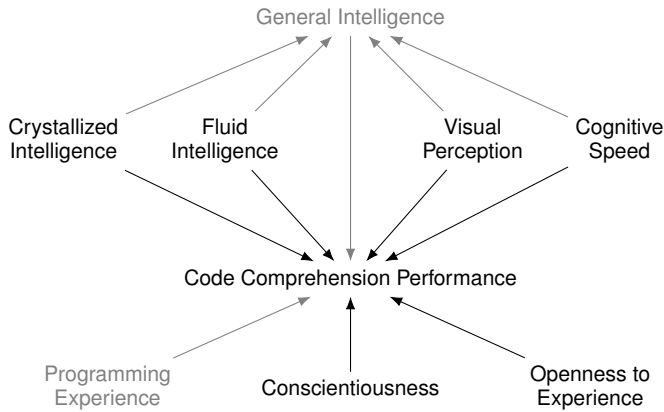

One way to improve causal inferences based on observational data are directed acyclic graphs (DAGs), which visually represent causal assumptions~\cite{Rohrer:2018:ThinkingClearly}.
Figure~\ref{fig:causal-diagram} shows the causal assumptions of the relevant variables in our study.
For example, we conjecture that a change in the variable crystallized intelligence causally contributes to a change in code comprehension performance.

We refer the interested reader to the work of Rohrer~\cite{Rohrer:2018:ThinkingClearly}, who explains in detail how such a graph can be used during study design to identify and eliminate spurious paths, for example by statistical control.
The whole approach is based on the assumption that the DAG captures the true underlying causal web, which we recognize is a very strong assumption.

We chose this method to argue with greater certainty for a causal relationship among the variables in the six hypotheses.
All authors of this paper were involved in the creation of the causal diagram during the study design.
The influence of known confounding parameters in program comprehension studies~\cite{Siegmund:2015:Confounding} was rigorously discussed and literature on known influences on intelligence and personality was consulted.

The only notable variable that we could not control via our study design was the motivation of the participants.
It may have an influence on the results of the intelligence test as well as on the results of the code comprehension tasks and thus represents a potential confounder.
To obtain a proxy for estimating participants' motivation to complete the test seriously, we used the number of requests for an individual participant's test results (about 40\%) as well as examined outliers in the data set.
The analysis provided grounds to assume that a considerable proportion of participants answered the tests seriously.
The descriptive statistics at the beginning of section~\ref{sec:results} show that the sample performed very well on the code comprehension tasks and the intelligence test scores are similar to those of the population.

Unless we have not controlled for any other factor that significantly influences both treatment and outcome, it is reasonable to assume that measured associations in hypothesis testing can be attributed to cause-effect relationships.
We nevertheless agree with Rohrer's concluding remark that \enquote{the most convincing causal conclusions will always be supported by multiple designs}~\cite{Rohrer:2018:ThinkingClearly} and call for further code comprehension studies on the given research questions to confirm or refute our model.

\subsection{Analysis Procedure}

We first used descriptive statistics to describe the sample. We employed the median (Mdn) and inter-quartile range (IQR) for ordinal data (self-assessed experience) and mean and standard deviation (SD) for interval data (code comprehension score). Also, intelligence and personality are considered interval data in psychology. Therefore, we did the same in our analysis.

Second, to control for the experience of the participants, we built linear models for each hypothesis to account for the factor affected in the hypothesis and the experience item with the strongest correlation to the score. We calculated standardised regression coefficients $\beta$ together with their 95\% confidence intervals (CI). We interpreted these intervals before using t-tests to calculate $t$-statistics and $p$-value as basis for statistical significance. To check the assumption of normality for the t-test, we conducted Shapiro-Wilk tests on the individual variables. For all but score and fluid intelligence, the test supported a normal distribution of the data. The QQ plots as well as the large sample size, however, make us confident that a t-test is still justified. 

Because we have multiple tests, we adjusted the $p$-values using the Holm-Bonferroni method~\cite{holm1979}. It is a standard method to adjust for multiple testing but is more powerful than the often used Bonferroni method. We compared the final $p$-values with the significance level $\alpha=.05$.

Third, for further exploratory analysis and to get a better understanding of the relationships of the various factors, we built further linear regression models. In particular, we analysed a ``complete'' linear model that contains all intelligence facets, the two personality traits also investigated in the hypotheses and all three experience items. This shows us the influence of all factors measured in the experiment. Furthermore, we included all measured factors (adding the other personality traits and the general intelligence to the ``complete'' model) and conducted step-wise regression on it to get to the best model for explaining the data. This gives us a smaller model with only the most important factors. For both models, we calculated the adjusted $R^2$ as a measure of how much of the variance in the data can be explained by the model, as well as standardised regression coefficients with their 95\% confidence intervals.

\section{Results\label{sec:results}}

\subsection{Descriptive Statistics}

We removed four participants from the data set because they provided data for the intelligence test and/or the personality test but not the comprehension test or the other way around. We have complete data for 117 participants,
further 17 participants without an intelligence test and one participant without a personality test.

\begin{figure}[htb]
    \centering
    \includegraphics[width=\columnwidth]{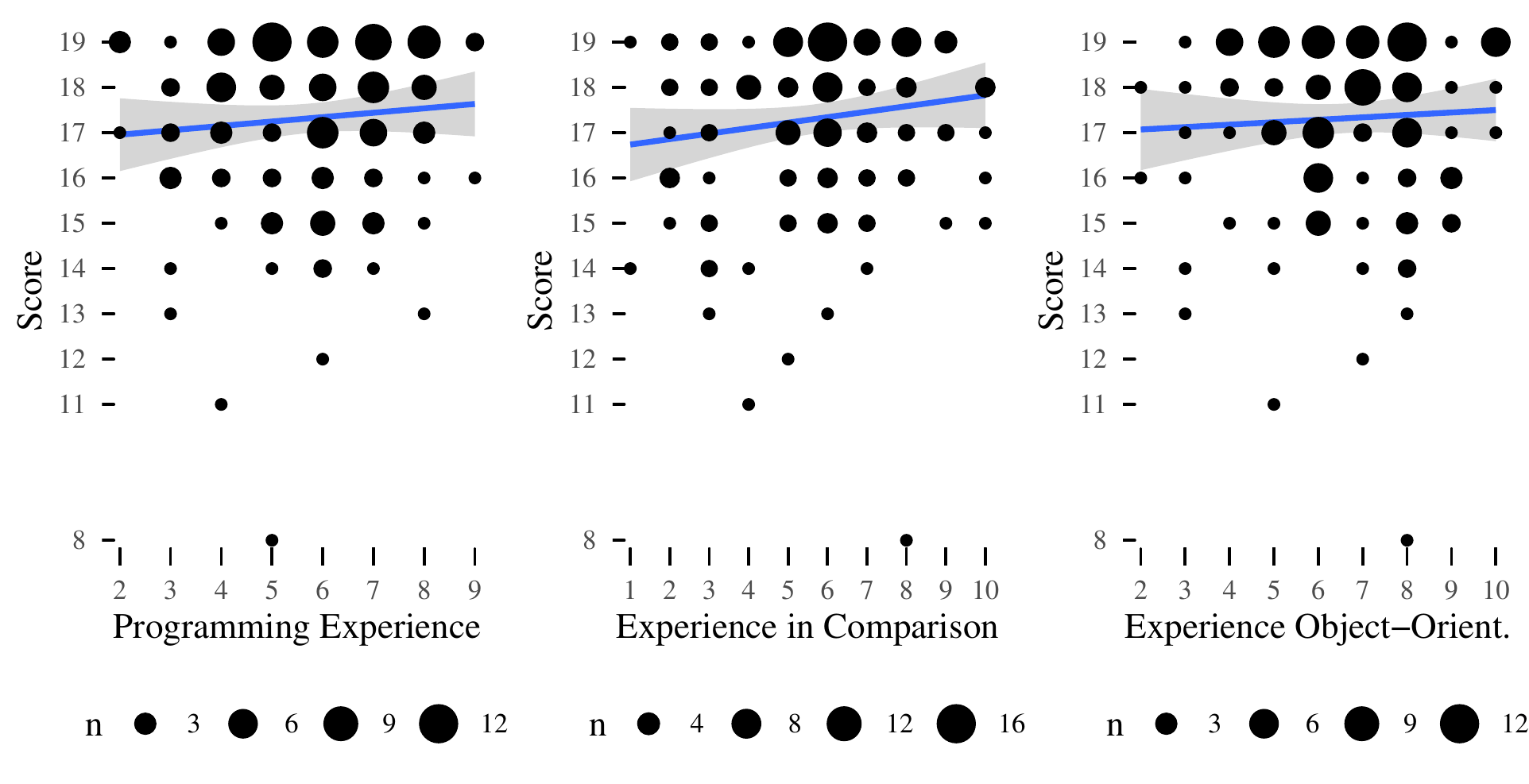}
    \caption{Overview of experience in relation to code comprehension scores}
    \label{fig:scatter-experience}
\end{figure}

The code comprehension performance as measured by the score concentrates strongly between 15 and 19, $M=17.3, \mbox{Min.}=8, \mbox{Max.}=19, \mathit{SD}=1.9$. The code comprehension score is depicted in relation to the three experience measures in Fig.~\ref{fig:scatter-experience}. The scatter plots show no
apparent strong relationships between the experience measures
and the code comprehension scores. The strongest relationship seems to be when the participants rated their experience in comparison to others. 

The three facets of experience, i.e., programming experience (PE), experience in comparison (EC) and experience in the object-oriented paradigm (EOO), have widely distributed results over the whole spectrum with a central tendency to slightly above the middle of the scale ($\mathit{Mdn}_\mathrm{PE}= 6, \mathit{IQR}_\mathrm{PE}=2, \mathit{Mdn}_\mathrm{EC}= 6, \mathit{IQR}_\mathrm{EC}=2, \mathit{Mdn}_\mathrm{EOO}= 7, \mathit{IQR}_\mathrm{EOO}=2$). Hence, we seem to have a well-balanced sample in terms of experience.

\begin{figure*}[htb]
    \centering
    \includegraphics[width=\textwidth]{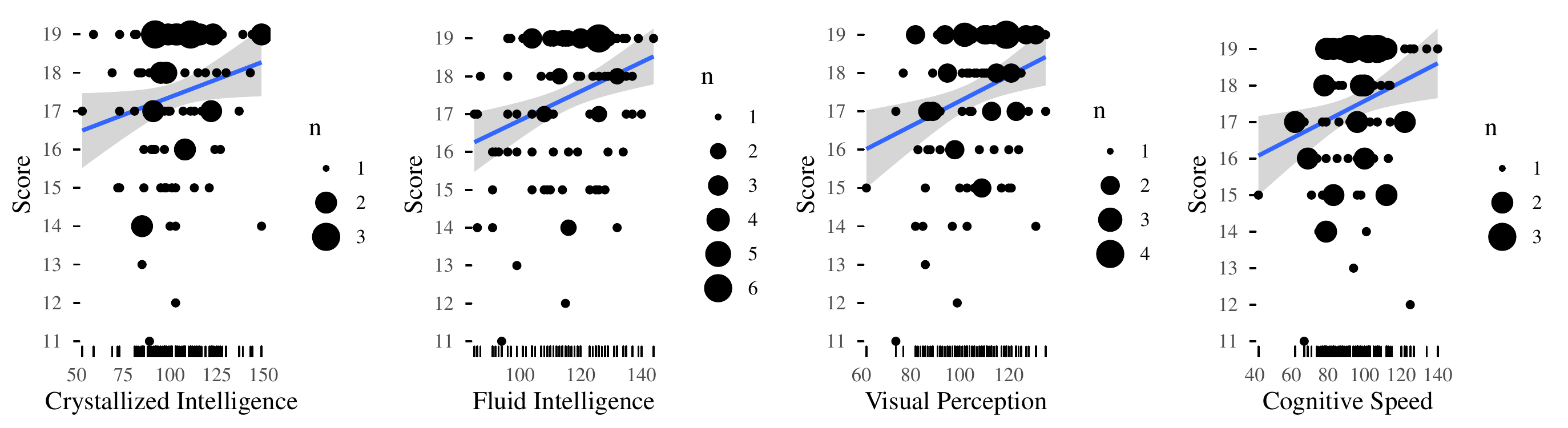}
    \caption{Overview of intelligence in relation to code comprehension scores}
    \label{fig:scatter-intelligence}
\end{figure*}

\begin{figure}[thb]
    \centering
    \includegraphics[width=\columnwidth]{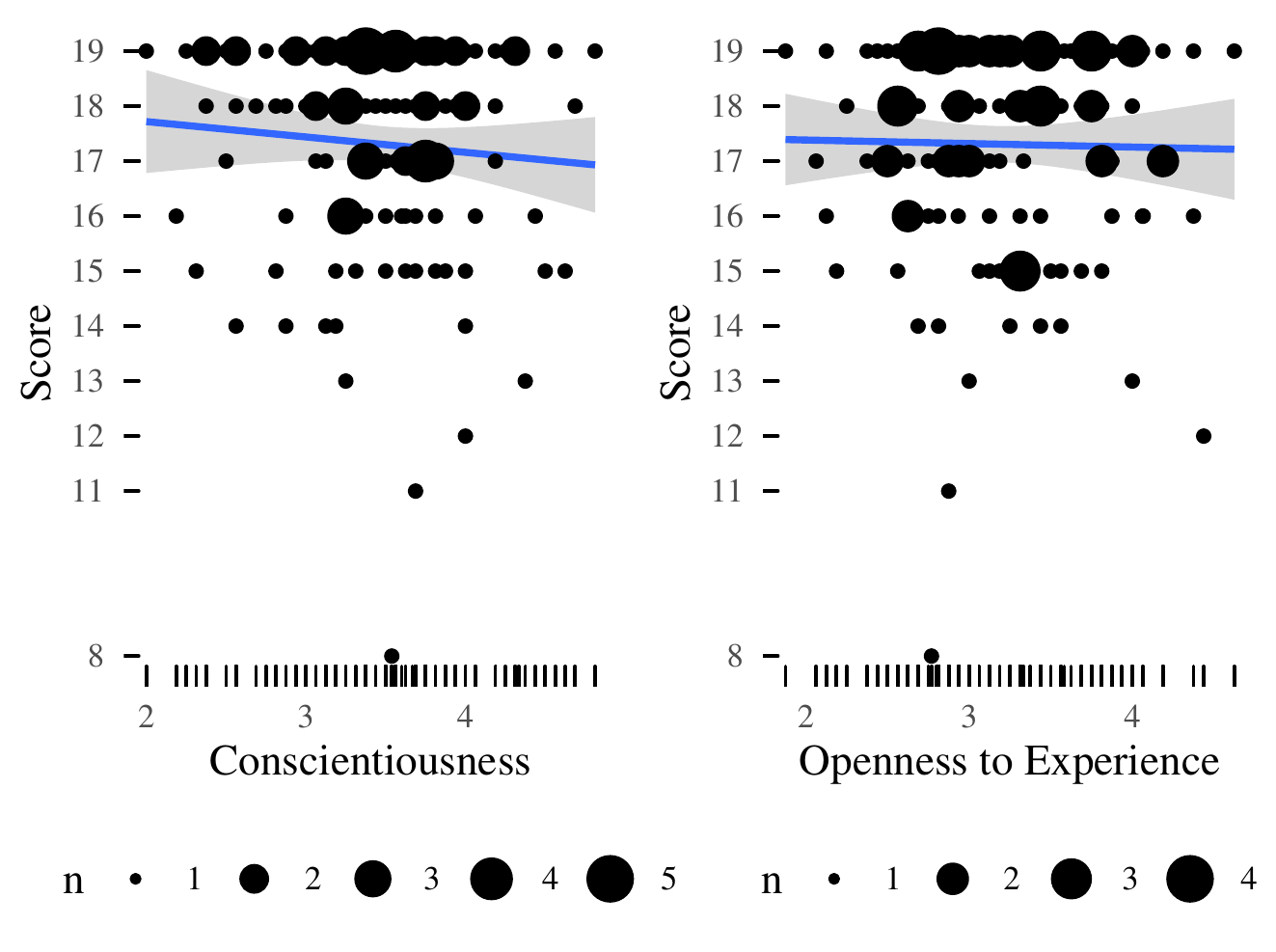}
    \caption{Overview of personality traits in relation to code comprehension scores}
    \label{fig:scatter-personality}
\end{figure}

In Fig.~\ref{fig:scatter-intelligence}, we see the relationships between the four different intelligence facets with the scores. As expected with intelligence tests, the results are distributed roughly around 100, but we also see rather low and rather high values in all of them ($M_{G_C}= 103.6, \mathit{SD}_{G_C}= 18.4, M_{G_F}= 115.6, \mathit{SD}_{G_F}= 13.9, M_{G_V}= 105.0, \mathit{SD}_{G_V}= 14.6, M_{G_S}= 94.5, \mathit{SD}_{G_S}= 16.7$). There appears to be a positive association between all the intelligence facets and the code comprehension score.

Finally, the data for the two personality traits conscientiousness and openness to experience is depicted in Fig.~\ref{fig:scatter-personality}. Both personality traits are strongly distributed over almost their whole spectrum. For conscientiousness, we have a mean result of 3.5 ($\mathit{SD}=0.6$). The mean for openness to experience is only slightly lower at 3.2 ($\mathit{SD}=0.6$). There appears to be no clear association of both traits with the code comprehension scores. Conscientiousness tends even to be slightly negatively related to the score.

\subsection{Hypothesis Tests}

The results of the hypothesis tests of all six hypotheses are summarized in Table~\ref{tab:hyp-tests}. We see clear positive standardised regression coefficients between each intelligence facet and the code comprehension scores when controlling for experience. They also constitute the effect size, which according to Cohen \cite{cohen1992power}, we can consider as small associations. The confidence intervals are all rather narrow so that the true coefficients will likely be a small positive association.

\begin{table}[htb]
\caption{Results of individual linear regression models}
\label{tab:hyp-tests}
\begin{tabular}{l r r r r r}
\toprule
Factor & $\beta$ & 95\% CI & $t$ & $p$ & Adj. $p$\\
\midrule
Crystal.~Int. & 0.188 & $[0.171$, $0.206]$ & 2.02 & .0460 & .1380\\
Fluid Int. & 0.299 & $[0.276$, $0.322]$ & 3.30 & .0013 & .0078\\
Visual Percep. & 0.261 & $[0.238$, $0.283]$ & 2.84 & .0054 & .0270\\
Cogn.~Speed & 0.247 & $[0.228$, $0.267]$ & 2.68 & .0084 & .0336\\
Conscientiousn. & -0.096 & $[-0.693$, $0.501]$ & -1.11 & .2710 & .5420 \\
Open.~to Exp. & -0.040 & $[-0.633$, $0.552]$ & -0.46 & .6440 & .6440\\
\bottomrule
\end{tabular}
\end{table}

For both personality tests, there are very small negative standardised regression coefficients. The confidence intervals, however, are very large and span from a medium negative to a medium positive association in both cases. Hence, from our data set, it is not clear if there is an effect at all and in which direction an effect would go.

To test the statistical significance, we use t-tests for which we report the test statistic $t$ as well as the corresponding $p$-value. Here again we see $p$-values for all intelligence facets lower than 5\% while both personality traits show $p$-values far greater than 5\%. To account for multiple testing, however, we adjusted the $p$-values using the Holm-Bonferroni method~\cite{holm1979}. The adjusted $p$-values give us almost the same picture. Only crystallized intelligence has now a $p$-value larger than 5\%. 

Hence, we have to reject the corresponding null hypotheses and have support for $H_2$, $H_3$ and $H_4$: There is a statistically significant, positive relationship between fluid intelligence,
visual perception and cognitive speed and code comprehension performance. Crystallized intelligence is positively related with comprehension performance, but this relationship is not statistically significant. There is no clear relationship between conscientiousness and openness to experience with comprehension performance.

\subsection{Exploratory Analysis}

Table~\ref{tab:linear-model} shows the results for the complete  linear model. Overall, the model explains only almost 10\% of the variance in the data (Adjusted $R^2 = .096$). The standardised regression coefficients show to some degree a different picture than the individual regression models for the hypothesis tests. Fluid intelligence and cognitive speed have slightly lower but similar regression coefficients. The coefficient for visual perception is close to zero, and the crystallized intelligence even has fully moved to a (very small) negative coefficient. The confidence intervals for these coefficients also support this, and are rather narrow. For the experience measures, surprisingly, only experience in comparison has a positive regression coefficient. Yet, for the other two measures, the confidence intervals show that the true values could be positive or negative. In any case, this also supports the usage of experience in comparison in the linear models for the hypotheses tests. For conscientiousness and openness to experience, the coefficients and confidence intervals are almost the same as for individual linear models.

\begin{table}[htb]
\caption{Complete linear model}
\label{tab:linear-model}
\centering\begin{tabular}{l r r r r}
\toprule
Predictor & $\beta$ & 95\% CI & $b$\\
\midrule
Intercept & 0.000 & $[-4.020$, $4.020]$ & 13.722\\
Crystallized Int. & -0.045 & $[-0.070$, $-0.020]$ & -0.004\\
Fluid Int. & 0.220& $[0.185$, $0.254]$ & 0.028 \\
Visual Perception & 0.057 & $[0.023$, $0.091]$ & 0.007 \\
Cogn.~Speed & 0.154 & $[0.129$, $0.178]$ & 0.016 \\
Exp.~in Comparison & 0.316 & $[0.079$, $0.553]$ & 0.278 \\
Experience OO & -0.061 & $[-0.308, 0.185]$ & -0.062 \\
Program.~Experience & -0.147 & $[-0.454, 0.162]$ & -0.163 \\
Conscientiousness & -0.103 & $[-0.697$, $0.491]$ & -0.327 \\
Openness to Exp. & -0.051 & $[-0.642$, $0.540]$ & -0.160 \\
\bottomrule
\end{tabular}
\end{table}

When looking at the non-standardised parameters, we see that the model has a large non-standardised intercept because most of the code comprehension scores are in the area above 13 points. 
All the intelligence facets have very small regression coefficients ($b$). Both personality traits have small to medium strength negative regression coefficients. Only experience in comparison has a considerable positive
regression coefficient. So for the direct prediction of the outcome of the results of our study, mostly experience and the personality traits would be important. 

As that model only explains 10\% of the variance, we wanted to further explore if we can find a better model. For that, we employed backward stepwise regression starting from a model that includes all measured variables from the experiment. It reduced the variables to the model shown in Table~\ref{tab:stepwise-linear-model}. This model explains more than 12\% of the variance (adjusted $R^2 = 0.122$). For this, we only need the three factors general intelligence, experience in comparison and conscientiousness.

\begin{table}[htb]
\caption{Linear model based on stepwise regression and all measured factors}
\label{tab:stepwise-linear-model}
\centering\begin{tabular}{l r r r r}
\toprule
Predictor & $\beta$ & 95\% CI & $b$ \\
\midrule
Intercept & 0 & $[-3.07, 3.07]$ & 14.8820 \\
General Intelligence & 0.305 & $[0.285, 0.325 ]$ & 0.033\\
Exp.~in Comparison & 0.174 & $[0.013, 0.335]$ & 0.152\\
Conscientiousness & -0.157 & $[-0.754, 0.439]$ & -0.511\\
\bottomrule
\end{tabular}
\end{table}

General intelligence has the strongest standardised regression coefficient with a narrow confidence interval in this model. The other two factors have roughly half the standardised regression coefficient, experience in comparison with a positive influence and conscientiousness with a negative influence. Both have wider confidence intervals. Especially the confidence interval of conscientiousness goes from a large negative to a medium positive value.

In summary, the exploratory analysis supports the results that fluid intelligence,
visual perception and cognitive speed are positively related to code comprehension performance.
It also supports the use of experience in comparison as the best subjective measure
for experience. Yet, it also shows that crystallized intelligence might actually be negatively
related to code comprehension performance when taking all factors into account. Furthermore, there
is indication that general intelligence might actually have the strongest relationship,
and personality traits could also play a role when considering interactions between
different factors.

\section{Discussion\label{sec:discussion}}

Literature in the field of code comprehension suggested that intelligence and personality might have an impact on code comprehension.
We were able to demonstrate such a relationship of correct answers to questions on code comprehension tasks to intelligence and personality traits.
The nature of these relationships must be considered separately for each factor, since some, such as fluid intelligence, appear to represent an independent influence on code comprehension performance, and  others, such as conscientiousness, appear to explain some of the variance only in combination with other factors.

Fluid intelligence ($\beta=0.299$), visual perception ($\beta=0.261$), and cognitive speed ($\beta=0.247$) showed significant association with code comprehension performance and can be preliminarily ranked in that order in their influence.
Cognitive speed could be of greater importance in scenarios where there is greater time pressure.
Crystallized intelligence showed the weakest standardised coefficient ($\beta=0.188$) and might be of lesser importance in code comprehension.

Our study is underpowered with 135 participants at a calculated optimal sample size of 198, which could mean that in a higher powered experiment, more effects could be detected by reducing the type 2 error. Crystallized intelligence might be statistically significant because it already has had an only positive CI in our experiment. Yet, the exploratory analysis pointed in the opposite direction: a negative association with comprehension performance. Hence, this relationship remains unclear from our experiment.
Similarly, a larger sample size might help to narrow the confidence intervals for the personality traits and make their influence clearer.

An exploratory analysis of age-adjusted general intelligence shows a moderate positive standardised coefficient of $\beta=0.312$ and at the same time general intelligence is the best predictor in the linear model based on stepwise regression~(Table~\ref{tab:stepwise-linear-model}) with a very small confidence interval.
This suggests that a combination of several high values for different intelligence facets is useful for the successful performance in code comprehension tasks.

We further noticed during the analysis that adding general intelligence to the linear model in Table~\ref{tab:linear-model} increases $R^2$ from 9.7\% to 12.4\%, thus explaining more of the variance in the data and supporting the common assumption that general intelligence is an independent factor that has its own impact on performance.

\vspace{0.2cm}
\noindent
\fbox{%
    \parbox{\columnwidth-0.25cm}{%
        \textbf{RQ1.} There is a positive relationship between individual intelligence factors and general intelligence with code comprehension performance. The combination of high values for different intelligence facets leads to high general intelligence, which in turn has the greatest predictive power for successful comprehension of code snippets.
    }%
}
\vspace{0.2cm}

The two personality traits examined for the hypothesis tests, i.e., conscientiousness and openness to experience, showed standardised coefficients close to 0 when controlling only for experience.
A further exploratory analysis shows similar results for agreeableness ($\beta=0.045$) and honesty/humility ($\beta=-0.061$). For emotionality ($\beta=0.161$) and extraversion ($\beta=-0.152$) we see at least small individual coefficients.

\vspace{0.2cm}
\noindent
\fbox{%
    \parbox{\columnwidth-0.25cm}{%
        \textbf{RQ2.} At first glance, personality does not appear to be a particularly interesting avenue for further research in the context of code comprehension performance.
        On their own, at least, individual personality traits do not have a significant impact on performance. However, we see that conscientiousness does play a predictive role in interaction with other factors. The linear models suggest that there may be other factors that interact with conscientiousness and may determine whether this personality trait has a positive or negative effect on code comprehension.
    }%
}
\vspace{0.2cm}

For the design of our study, we used a causal diagram~(Fig.~\ref{fig:causal-diagram}) that helped us identify spurious paths and covariates, for example.
One such covariate is programming experience, and it was confirmed in our data to have a significant impact on performance in code understanding.
Of the three experience measures proposed in~\cite{Siegmund:2014:MeasuringExp}, experience in comparison is best suited to predict performance, although the CI is wide and thus the actual impact unclear. At least the CI is all positive, and in a homogeneous sample it might already be sufficient to query this one item as a proxy for the experience of a participant.

Siegmund et al.'s study~\cite{Siegmund:2014:MeasuringExp} is comparable in terms of participant and task characteristics. They had a homogeneous student sample of 128 participants whose task consisted largely of having to mentally simulate code to determine the output of that code. Moreover, code comprehension was measured via correctness in these tasks. Their final recommendation that in a student population, experience in comparison might be sufficient to reliably measure programming experience is supported by our data. Our data differ in that the experience measures \textit{programming experience} and \textit{experience with object-oriented languages} are not positive factors influencing performance. We agree with their view that additional experiments are needed to construct a valid experience measure.

What is apparent from our results is that our causal diagram may not be complete or individual connections assumed to be causal relationships may not be causal in nature (or at least cannot be considered isolated from other factors).
Our optimized linear model provided in Table~\ref{tab:stepwise-linear-model} explains only a small portion of the variance in the code comprehension scores, i.e., intelligence, experience and conscientiousness do not seem to be sufficient to predict code comprehension performance as we have operationalized it.
We see two possible explanations for this.

First, it may be that the specific way of measuring code comprehension affects the strength of the predictors studied.
For example, the selection of our tasks and how we scored them led to low variance in code comprehension scores, which in turn tends to lead to lower regression coefficients in general.
We elaborate on the relevance of task design and construct measure in the following subsection.

Second, code comprehension could be a construct that includes skills that are not captured by intelligence tests.
One might even assume a missing individual characteristic of a developer in the causal diagram, which serves as a predictor for code comprehension proficiency.
This, in turn, argues for measuring code comprehension via experimental procedures and tasks tailored to code comprehension, as has been done to date, rather than being replaced by existing psychometric tests designed to measure related constructs such as intelligence.

\subsection{Limitations\label{sec:limitations}}

The results of this study should be seen in the light of some limitations.

First, we invited a convenience sample of students in their second year, so the results should only be generalized to more experienced developers with caution.
While we have a well-balanced sample in terms of self-reported programming experience, it is clear that second-year students on average have less experience than students in higher semesters or even graduates, and such increased experience may lead to different results in the code comprehension tasks.
In terms of the distribution of intelligence, our sample appears to be representative of university students, with slightly higher mean values for fluid intelligence and slightly lower for cognitive speed in our sample compared to the normalization sample~\cite{lps2}.
However, we lack data to assess representativeness for all developers, regardless of their educational background.

Given the expected sample characteristics, we limited our selection of tasks and code snippets to those that were of easy to moderate difficulty and thus tend to be in line with comparable code comprehension studies~\cite{Baron:2020:Empirical}.
Unfortunately, in our study, this resulted in the majority of the scores achieved in the code comprehension tasks being concentrated in a range close to the maximum score.
Thus, while we were able to distinguish between the average participant and low performers, we were likely not able to identify nuances in the performance of individual participants sufficiently well.
This in turn explains the low regression coefficients in the linear model.
We recommend repeating the experiment with a wider range of code snippets of varying difficulty or changing the operationalization of code comprehension performance.

Related to this, it is in the nature of our study that we consider code comprehension as an isolated process that our participants had to go through on their own.
While such a controlled setting is good for identifying individual influences on performance with satisfying internal validity, we are aware that in practice, for example, code comprehension is accompanied by other activities or even that developers sometimes read code together with their peers.
We are certain that studies in more realistic settings would provide many additional valuable insights into the effects of conscientiousness and individual intelligence factors on code comprehension performance and behaviour.
We consider our findings to be a valuable starting point that provides evidence for the influence of two constructs on the isolated cognitive process of code comprehension.
Additional studies, including those of a qualitative nature, are needed to shed light on our research questions from other interesting angles.

Code comprehension as a construct has been measured solely by the correctness of answers to comprehension questions.
While there is no validated measure of code comprehension so far, we suspect that correctness alone does not capture all facets of code comprehension.
It would therefore be interesting to see which additional information we would gain if, for example, comprehension efficiency (see, e.g.,~\cite{Scalabrino:2019:Automatically,Wyrich:2020:Mind}) or cognitive load~\cite{Fakhoury:2018:Objective} were instead used as proxies for code comprehension.
Due to our study design with a large number of participants, in presence and synchronous execution, we were limited in this respect, but aim for complementary studies in the future.
Apart from that, we found that conducting our study synchronously with a three-digit number of participants worked smoothly.
This type of code comprehension measurement scales very well, provided that large rooms are available to the study leaders.

Compared to the measurement of code comprehension, research on the measurement of personality and intelligence is more advanced.
We were able to use validated questionnaires for their measurement, but like code comprehension, they are latent variables that only approximate what currently corresponds to (parts of) our definition of them.
Accordingly, we would like to note that our conclusions on the influences of personality and intelligence in this work are to be considered under the assumptions and definitions that the respective tests establish for these constructs.
In this specific case, we do not consider this to be a significant limitation, since the used instruments are well-established questionnaires whose results can usually be compared with those of other studies on personality and intelligence.

\subsection{Implications}

Since several intelligence factors correlate significantly with developers' code comprehension performance, researchers have so far probably been correct in their assumption that not controlling intelligence in their study design is a potential threat to validity~\cite{Siegmund:2015:Confounding}.
For example, if one of two experimental groups has a significantly higher mean intelligence value, this will have a positive effect on the code comprehension performance of that same group and, as is common for confounding variables, may lead to false conclusions about the influence of the independent variable that is actually being measured.
This situation remains even if our assumptions about the causal relationship are refuted in the future.
The measured correlations remain valid and explain some of the variance in code comprehension performance.

However, these findings do not mean that previous research on code comprehension performance is invalidated, but only that intelligence may have had an impact on the measured code comprehension data in some cases.
Moreover, for now, the limitations of the transferability of our results to studies that also have a sample of university students apply.

As for the influence of personality traits, we see great potential for more in-depth studies examining the specific nature of the relationship to code comprehension and the interplay with other factors. Conscientiousness seems to play a role in predicting code comprehension performance, but apparently only in combination with other factors and probably with some we did not measure in our study.

Consequently, what should be considered in the design of future code comprehension studies? 
Code comprehension studies in which intelligence and personality might play a role and which do not control for these constructs by design risk bias in their results.
If intelligence and personality traits are not explicitly controlled by, e.g., matching or post hoc analysis, because it would not be feasible to have every participant take an intelligence and personality test, then other control techniques should be used.
One mentioned by Siegmund and Schumann~\cite{Siegmund:2015:Confounding} is randomziation and the notable advantage of this technique is that one may also control for additional, possibly even unknown extraneous variables.\\
Alternatives we see are first the use of validated, reduced IQ or personality tests, which take only a fraction of the time of full-scale tests, but at the same time are a useful approximation.
Second, we suggest a more solid discussion of threats to validity based on data about the influence of potential confounders.
For example, in a quantitative study of the influence of syntax highlighting on code comprehension in which intelligence was not controlled, if the effect size is large enough, it can at least be argued that the measured influence of syntax highlighting is larger than that expected in groups with significantly different intelligence distributions.

Independent of controlling for a potential confounding factor, we consider it to be necessary and at the same time equally interesting for a study to measure intelligence and personality to examine their individual influences in the context of a specific research question and thus enrich our more abstract results.
While we consider intelligence and personality as potential confounding variables in the context of this study, they are in the end also given individual characteristics of each developer, and if we could better understand their influence on the work of those same developers, we can support developers outside of controlled experiments through, for example, appropriately customizable tools and consideration of individual capabilities.
Accordingly, future work can build on our findings to explore their concrete consequences in practice and, in the long run, to develop measures, if necessary, to counteract potentially negative consequences of inequalities in given individual characteristics.

Finally, the present study has shown that a potential influence of intelligence discussed in the literature not only actually has an impact on performance, but also provides insight into the direction and strength of individual facets. We appreciate that it has now become the norm to discuss threats to validity in code comprehension studies. Backing these discussions up with evidence in the future should be the next step in the maturation of the research field and for this it needs further studies that identify new confounders or confirm or refute assumed ones.

\section{Conclusion\label{sec:conclusion}}

The list of suspected confounding parameters on program comprehension is long~\cite{Siegmund:2015:Confounding}.
Taking them all into account in the design of valid studies remains a challenge.
It is therefore essential that we obtain certainty about the extent and the nature of the relationship of each parameter to code comprehension through empirical studies, for example, to better evaluate alternative study designs.

We investigated the influence of intelligence and personality on code comprehension performance in a study with 135 university students.
While personality traits showed no association with performance on their own, we found significant small to moderate positive association between code comprehension performance and the intelligence factors fluid intelligence, visual perception and cognitive speed.
We found a weak relationship of performance to crystallized intelligence, which was not statistically significant.
An exploratory investigation further showed a moderate positive relationship of performance with general intelligence and that it is the variable with the greatest predictive power in our linear model.

Given our sample size, the measured regression coefficients are large enough for intelligence to be a noteworthy potential confounding variable, especially from a researcher's perspective.
The results indicate that the control of intelligence in code comprehension experiments is necessary for valid conclusions from obtained study data.
We draw a similar conclusion for the personality trait conscientiousness, although the specific nature of the relationship to code comprehension requires further research.
At the same time, the results should be considered in light of the selected tasks, sample characteristics, and code comprehension measures, which is why we encourage further studies on the relationships between intelligence and personality with code comprehension performance to enhance our understanding of these relationships.
The more in-depth investigation of potential confounding parameters, be it intelligence, personality or any other variable, will eventually lead to more confidence in the validity of code comprehension studies.

\section{Data Availability\label{sec:data}}

We disclose code snippets, task sheets with comprehension questions, anonymized raw data, and the R script for the analysis openly~\cite{zenodo:dataset}. Please note that the raw data does not contain the complete data set, as we only make the data of those participants public that explicitly agreed to it.

\section*{Acknowledgements\label{sec:acks}}

We are grateful to our participants for taking part in our study. We thank three anonymous reviewers for their constructive, insightful, and encouraging feedback.

\balance

\bibliographystyle{IEEEtran}
\bibliography{bibliography}

\begin{IEEEbiographynophoto}{Stefan Wagner}
is full professor for empirical software engineering and director of the Institute of Software Engineering at the University of Stuttgart, Germany. He studied computer science in Augsburg and Edinburgh and received a doctoral degree from the Technical University of Munich. His research interests include software quality, requirements engineering, agile development, DevOps and software engineering for AI-based systems. He is a senior member of IEEE and a member of ACM and the German GI.
\end{IEEEbiographynophoto}

\begin{IEEEbiographynophoto}{Marvin Wyrich}
is working toward a PhD degree in computer science at the University of Stuttgart, Germany, where he is part of the empirical software engineering research group since 2018. His research interests include empirical and behavioural software engineering, with a focus on the design of reliable and valid program comprehension studies.
\end{IEEEbiographynophoto}

\end{document}